\documentclass[twocolumn,pra,floatfix]{revtex4}
\usepackage[T1]{fontenc}
\usepackage[latin1]{inputenc}
\usepackage{graphicx}
\usepackage{color}
\usepackage{amsmath}
\usepackage{amssymb}
\usepackage{amsthm}
\usepackage{bbold}
\usepackage[export]{adjustbox}
\usepackage[capitalize]{cleveref}
\usepackage{makecell}
\usepackage[font=footnotesize]{subfig}
\usepackage[section]{placeins}
\usepackage[toc,page]{appendix}

\def\bf#1{\mathbf{#1}}
\newcommand\Tstrut{\rule{0pt}{8ex}}         
\newcommand\Bstrut{\rule[-6.5ex]{0pt}{0pt}}   
\crefname{appsec}{Appendix}{Appendices}
\newtheorem{theorem}{Theorem}

\begin{document}

\title{Chiral Maxwell waves in continuous media from Berry monopoles}
\author{M. Marciani and P. Delplace}
\affiliation{Univ Lyon, Ens de Lyon, Univ Claude Bernard, CNRS, Laboratoire de Physique, F-69342 Lyon, France}

\begin{abstract}

    We propose a method to predict the existence of topologically protected electromagnetic chiral modes between continuous media described by Maxwell's equations. The number and character of these modes is related to topological charges (Berry monopoles) in parameter space. Unlike the approaches proposed so far, our description does not require a regularization parameter at large $\mathbf{k}$ nor the materials to be topological on their own. The predictions of the theory are confirmed by additional numerical simulations of interfaces of gyrotropic media.
\end{abstract}

\maketitle

\section{Introduction}

In the last decades, the inclusion of topological analysis in condensed matter systems has lead to a more comprehensive understanding of their properties  \cite{Tho82,Ber84} and inspired applications for a technological advance beyond the realm of quantum electronic systems  \cite{Naya2008,Qi2011}, most notably in classical waves physics such as photonics \cite{Ozaw2019}, mechanics \cite{Kane2014, Nas2015,Hube2016} or acoustics \cite{Yang2015,Zha2018}

 A remarkable consequence of a non-trivial topology in wave physics is the presence of uni-directional (chiral) eigenmodes that perfectly propagate along interfaces between two materials  \cite{Wang2009} despite the presence of defects or weak disorder. The key mathematical object to determine the number of such modes in two-dimensional setups is the so-called (first) Chern number. Notably, this quantity can be computed only in systems where the momentum space of the bulk  materials is compact, mainly limiting the applicability of this tool to crystals, where the natural base space is the two-dimensional (2D) Brillouin zone that is a torus  \cite{Prod2016}. As a consequence the topological analysis of chiral waves in well-known continuous models appearing naturally in the context of optics, geophysics, mechanics and fluid mechanics has been put aside for a long time, being introduced only recently \cite{Sil2015,vanM2018,Sil2019b,Sou2019,Tau2019b}. Actually, the unbounded momentum space of such models is equivalent to a compact one if the parameters involved in the partial differential equations satisfy certain conditions in the large-momentum limit  \cite{Vol1988, Tau2019}. Though simple and powerful, this approach suffers of being not obviously applicable to real materials, since most continuous approximations inevitably break down in the large momenta limit.

 Here we suggest a different and more universal route where the emergence of interface chiral optical waves is predicted from the existence of degeneracy points in parameter space. These points behave as topological charges, or Berry monopoles, whose flux through a close surface surrounding them is also a Chern number. In that sense, the topological origin of interface chiral states is encoded into a local quantity that can be seen as a topological defect in parameter space, and thus does not require any additional regularization at large wavenumber as it is usually done in optical continua \cite{Sil2015,vanM2018}. This approach is therefore particularly suitable to address topological properties in continuum systems in general, and was recently used to reinterpret \cite{Del2017} and predict \cite{Per2019} topological fluid waves in the geo/astro-physical context. 
Adapting this method rigorously to the realm of electromagnetism in local and non-local continuous media, we apply the theory to predict the existence of chiral optical waves at the interface between metals, gyro-electric and gyro-magnetic materials.\\

The paper is organized as follow~: we briefly introduce the macroscopic Maxwell's equations without external sources and discuss their reduction to a (2+1)D setting (\cref{sec: max_intro}). Through the mathematical notion of fiber bundles, we present in \cref{sec: top_and_bb} what we refer to as the "standard bulk/boundary correspondence formula", widely employed in literature and applicable when the wavenumbers are defined in a compact space (e.g. in crystals and some non-local optical continua). The main contribution of our paper is presented in \cref{sec: spec_flow} where we generalize the formula to systems whose wavenumbers are not bounded, enabling a treatment of generic continua media. In \cref{sec: anal_num}, we test our theory on various  gyrotropic media. We first illustrates it on analytically treatable models such as the formal analog to the standard shallow water model encountered in geophysical flows, that is obtained here in a low frequency limit. Then, more complicated interfaces are analyzed beyond this approximation and treated numerically. The numerical finding of chiral Maxwell waves is in agreement with our topological method. In \cref{sec: sw_dirac_interf} we discuss a toy model beyond the framework of Maxwell equations which shows in a striking way how our theory can address situations that are not treatable within the other approaches proposed so far. Finally, we conclude in \cref{sec: concl}.

\section{ Homogeneous Macroscopic Maxwell's Equations}  \label{sec: max_intro}

The macroscopic Maxwell's equations govern the behavior of the macroscopic averages of the electromagnetic fields. In particular, setting charges and currents to zero, the equations describe the 3+1D dynamics of the four fields $\bf D,\bf E,\bf B,\bf H$:  
\begin{align} 
    \nabla \times \bf E &= -\frac{\partial \bf B}{\partial t} \quad\quad&     \nabla \times \bf H &=  \frac{\partial \bf D}{\partial t} \label{MaxSpace1}\\
    \nabla \cdot  \bf D &= 0 \quad\quad&     \nabla \cdot  \bf B &= 0. \label{MaxSpace3}
\end{align}
This set of partial differential equations are complemented by material-dependent constitutive equations relating algebraically the fields. We assume they are of the most general form that accommodates homogeneous and linear responses including magneto-electric effects
 \begin{align} \label{M_conv}
     \begin{pmatrix}
     \bf D \\
     \bf B \end{pmatrix} = M\, \circ\,
     \begin{pmatrix}
     \bf E \\
     \bf H \end{pmatrix} \quad \text{where}\ M := \begin{pmatrix}
         \varepsilon   &  \xi \\
         \zeta         &  \mu 
           \end{pmatrix} \, .
 \end{align}
Here, $\varepsilon$ and $\mu$ are respectively the permittivity and permeability $3 \times 3$ tensors while $\xi$ and $\zeta$ are the magneto-electric ones \cite{Ast1961,Pru2016}. The composition symbol $\circ$ denotes convolution in both space, time and the vector indexes. Here and later, to simplify the notation, we set $c=\epsilon_0=\mu_0=1$.
 
We will assume that the materials are constant in time. After Fourier transforming and going to the frequency domain it is apparent that, at finite frequency, \cref{MaxSpace3} are implied by \cref{MaxSpace1}. Since we are interested in finite frequency features, the divergence equations will not be relevant and we can neglect them in the reminder of the paper.

\cref{MaxSpace1} can be rewritten in a more compact way as a single "Maxwell operator" $L$ acting upon the full electromagnetic field $\bf V = \left( \bf E, \bf H \right)^T$~:
 \begin{align}
    L_{\omega}[\bf x,-i\bf \nabla] \cdot \bf V &= \left(iR[-i\bf \nabla] + \omega \,M_{\omega}[\bf x, -i\bf \nabla] \right) \cdot \bf V = 0   \label{Max1_operator}
\end{align}
where
 \begin{align}
    R[-i\bf \nabla] = \begin{pmatrix}
                        0         &  -i\,\bf S \cdot \bf \nabla \\
           i\,\bf S \cdot \bf \nabla  &            0 
           \label{eq:rot}
           \end{pmatrix}
\end{align}
is the $6\times 6$ matrix associated to the curl part of the equations and does not depend on $\omega$; here $({S^i})_{\alpha\beta} = i\epsilon^{i\alpha\beta}$ is formed out of the totally antisymmetric rank 3-tensor and is an element of a spin-1 $SU(2)$ algebra \cite{vanM2018,Bli2019}. For more details about the notations of operators, see \cref{app:symbol}.

In non dispersive materials, i.e. where $M$ does not depend on $\omega$, finding the frequencies supported by the material, consists in solving a generalized eigenvalue problem of form $i \,R \,\bf V = \omega M \,\bf V $. However, when the material is dispersive the dependence in $\omega$ cannot be singled out and the frequencies $\omega$ are determined in a more implicit "self-consistent way".
 
For homogeneous  materials $M$ does not depend on $\bf x$ and we can further Fourier transform in space ($-i \,\nabla \rightarrow \bf k$). The operator $L_{\omega}(-i \bf \nabla) \rightarrow L_{\omega,\bf k}$ becomes a matrix, the field $\bf V(\bf x)\rightarrow \bf V_{\bf k}$ a vector and the equation is reduced to an algebraic one:
 \begin{align}
    L_{\omega,\bf k} \cdot \bf V_{\bf k} &= (iR_{\bf k} + \omega \,M_{\omega, \bf k}) \cdot \bf V_{\bf k} = 0   \label{Max1}
\end{align}
 
We assume the matrix $M$ to be Hermitian (lossless materials condition) and to be real when expressed in spatial coordinates, hence $M_{\omega,\bf k } = M^\dagger_{\omega,\bf k } = M^*_{-\omega,-\bf k}$. Moreover, as $M$ describes causal responses, it is analytical in the upper half complex $\omega$-plane. We also take that the residues of $M$ are semi-positive definite, which guarantees the semi-positiveness of the energy fields. Few important properties follow: the eigenfrequencies $w_{\bf k}$ of \cref{Max1} are real \cite{Rag2008}; $L_{\omega, \bf k}=-L_{-\omega,- \bf k}^*$; each solution $\left(w_{\bf k}, f_{\omega_{\bf k},\bf k}\right)$ at finite frequency pairs up with a solution at opposite energy and momentum $(-w_{\bf k},\bf V_{-\omega_{\bf k},-\bf k}^*)$. The latter conditions identify a symmetry which we will refer to as particle-hole symmetry throughout the paper, by formal analogy with condensed matter. In an inhomogeneous system we can do similar assumptions on the self-adjoint (matrix-valued) operator $M_{\omega}[\bf x, -i\bf \nabla]$ and analogous consequences for $L_{\omega}[\bf x, -i\bf \nabla]$ follow.

We are interested in interface modes in (effectively) 2D materials in the $(x,y)$ plane. Usually there are two ways to define a 2D problem: the first one is to physically confine the fields along the plane with two metallic plates \cite{Jac1999}, and the second is to deal with a 3D system but assuming invariance along the transverse $z$ direction ($k_z=0$ in \cref{Max1}). For the sake of simplicity, we shall opt for the second one. For the discussion of examples, later we shall consider uniaxial responses of gyrotropic materials concerning either the permittivity tensor (gyro-electric effect) or the permeability tensor (gyro-magnetic effect) with $\hat z$ as the principal axis. In these cases, the $6\times 6$ set of Maxwell equations simplifies into two uncoupled sets of $3\times 3$ equations operating on $(E_x, E_y, H_z)$ and $(H_x, H_y, E_z)$ that are respectively referred to as transverse magnetic (TM) and transverse electric (TE) modes. 

\section{Topology and bulk-boundary correspondence} \label{sec: top_and_bb}

2D materials with broken time-reversal symmetry ($M \neq M^T$) can support the socalled \emph{ topological chiral states}. These are modes flowing unidirectionally along edges of materials or, more generically, interfaces and are robust to disorder. A remarkable mathematical property called \emph{bulk-boundary correspondence} connects the information of the number and direction of chiral modes to a topological invariant of algebraic equations avoiding a direct calculation of the eigenfunctions of the inhomogeneous Maxwell's equations. 

 The basic tool for a bulk-boundary correspondence is the vector bundle associated to each bulk band. To define them, we solve \cref{Max1}, compute the bands dispersion $w^{(i)}_k$ and the associated eigenvectors $\bf V^{(i)}_k$. Then for each (w.l.o.g. non-degenerate) band we can define a fiber bundle, that consists in the collection of the projectors onto the eigenvectors of a band $P_k = \bf V_k\, \bf V_k^\dagger$ (called fibers) parametrized by $(k_x,k_y)$, elements of $\mathbb{R}^2$ (called the base space, $\mathcal{M}$).

For a generic fiber bundle we shall consider the scalar quantity for each band: 
\begin{equation} \label{ch_num}
    C = \frac{1}{2 \pi i}\,\int_\mathcal{M} \mathrm{d} c_1 \, \mathrm{d} c_2 \,\mathrm{Tr} \,P \,\left(\partial_1 P\, \partial_2 P - \partial_2 P \partial_1 P \right)
\end{equation}
where $c_{1,2}$ are two generic coordinates on the base space and $\partial_{c_j}$ the associated partial derivative.

When $\mathcal{M}$ is compact, $C$ has the remarkable property to be an homotopy invariant integer, and is called the (first) Chern number.
The Chern number can be used to quantify a number of physical effects. In particular, the number of unidirectionnal (or chiral) states $N^{\sigma}_{chiral}$ localized at an interface between two gapped materials, and whose frequency migrate toward a band $\sigma$ as the momentum along the interface goes from lower to higher values, is computable via a so-called bulk/boundary correspondence \cite{Gra2013,Prod2016}:
\begin{equation} \label{bb_corresp1}
    N^\sigma_{chiral} = C^\sigma_{right} - C^\sigma_{left} 
 \end{equation}
where $C_{left/right}$ is the band Chern numbers computed from the bulk response of the left or right material relative to the band $\sigma$. Negative values of $N^\sigma_{chiral}$ mean opposite flow direction.

\cref{bb_corresp1} can be applied in lattices, such as photonic crystals, where the first Brillouin zone form a compact base space. On the contrary, the formula is not directly applicable in continuum systems like ours, as $\mathcal{M} = \mathbb{R}^2$ is non-compact, therefore $C$ is allowed to take non-integer values\footnote{In Ref. \onlinecite{Sil2015} it has been shown an instance of non integer $C$ for a band of a gyrotropic material. However the band taken into account is degenerate with other bands at $\bf k=0$ thus the Chern number of that band alone does not exist. An good and clean instance of non-integer $C$ is the massive spin $1/2$ Dirac Hamiltonian in the continuum. One can easily compute $C=1/2$ in this case.}. To define an integer Chern number for a band in continuum problems, further assumptions on the behaviour of $M$ at large $|\bf k|$ are to be made to make the base space $\mathcal{M}$ effectively compact \cite{Vol1988, Sil2015,vanM2018, Sou2019, Tau2019}.
However, in the large $|\bf k|$ limit the small scale structure of the material, being either inhomogeneous or crystalline, comes out. Therefore a microscopic approach beyond the macroscopic Maxwell's equations is demanded, implemented either with field sources ($\rho$ and $\bf J$) or Bloch theory respectively \cite{vanM2018b}. 
For that reason, we propose to follow a different strategy to predict the existence of interface chiral modes.\\

\section{Spectral flow from Berry Monopoles} \label{sec: spec_flow}
The approach we follow essentially consists in describing a continuous interpolation between two media where the gap in the ''bulk'' spectrum closes at some point during the interpolation. The quantized Berry flux emanating from this degeneracy point is a first Chern number that fixes the number of chiral modes trapped at the interface. The advantage of this method is that it does not require the bulk bands of the continuous materials at each side of the interface to have (well-defined) non-zero Chern number. In spirit, this approach is very similar to a 2D version of the celebrated Jackiw-Rebbi model where the mass term in the Dirac equation changes sign along a direction  \cite{Jac1981, Teo2010}. An optical analog of this model, where the anisotropic mass term is played by an anisotropic Faraday effect, was already derived by Raghu and Haldane in their pioneering paper on topological photonics  \cite{Rag2008}. However, this model was derived as an effective description in the vicinity of a Dirac point emerging in the full band structure of the photonic honeycomb crystal, so that the topological properties were derived by means of the compact Brillouin zone.

\subsubsection{Interface model} 
To be definite, suppose we have two semi-infinite 2D materials in contact along the $y$-axis (i.e. at $x=0$) and characterized by two different bulk responses $M_1$ and $M_2$ respectively. We assume that the interface is defined by a \textit{smooth} change in an interval of width $x_0$ of an extrinsic parameter (e.g. an applied magnetic field) or intrinsic one (e.g. the permittivity tensor).
As a consequence, Maxwell's equations become inhomogeneous along $x$ and one should work with \cref{Max1_operator}. The action of the response matrix can be written through its kernel in the following form:
\begin{equation} \label{M_subst}
M_{\omega}[\bf x,-i\bf \nabla] \cdot \bf V  (\bf x)= \int_{\mathbb R^2} \mathrm{d} \bf x' \; M_{\omega}( x, \bf x - \bf x')\cdot \bf V (\bf x')
\end{equation}
with $M_{\omega}( x, \bf x - \bf x')$ being a distribution converging to $M_{1,2;\omega}(\bf x - \bf x')$ for $x\gg x_0$ and $x\ll - x_0$ respectively.

For instance, in the following sections we will consider the linear interpolation (we drop the dependence from $\bf x - \bf x'$):
\begin{equation} \label{M_interp}
 M(x) = \frac{M_1\!\! +\!\! M_2}{2} + \tau(x)\frac{M_1 \!\! -\! \! M_2}{2} , \;\;\tau(x) = \tanh(x/x_0) 
\end{equation}

\subsubsection{Symbol and spectral flow formula}

 To study the qualitative (topological) properties of inhomogeneous Maxwell's equation \cref{Max1_operator}, the modern analysis of pseudo-differential equations offers us a precious tool called the "symbol" of an operator \cite{Tay1996,Zwo2012}. It is a standard result that a non-trivial topology of the symbol gives information on the spectral flow of the associated operator. While we leave the details of its rigorous definition in \cref{app:symbol}, we introduce it here in a more sloppy but intuitive way. 
 
 In essence, to define the symbol associated to the Maxwell's equations operator we can regard $x$, the parameter describing the interface, as an external parameter independent from $i\partial_x$ (or $x - x'$ if we use a kernel representation, cf. \cref{M_subst}) in the Maxwell's equations. In doing so the equations become fully homogeneous again and we can Fourier transform in space similarly to what was done to get \cref{Max2}. Thus the original operator $L_{\omega}[\bf x,-i\bf \nabla]$ gets represented by a matrix $L_{\omega,\bf k, x}$, which is its socalled associated (standard) symbol.
 
\begin{figure}
\includegraphics[width=0.45\textwidth]{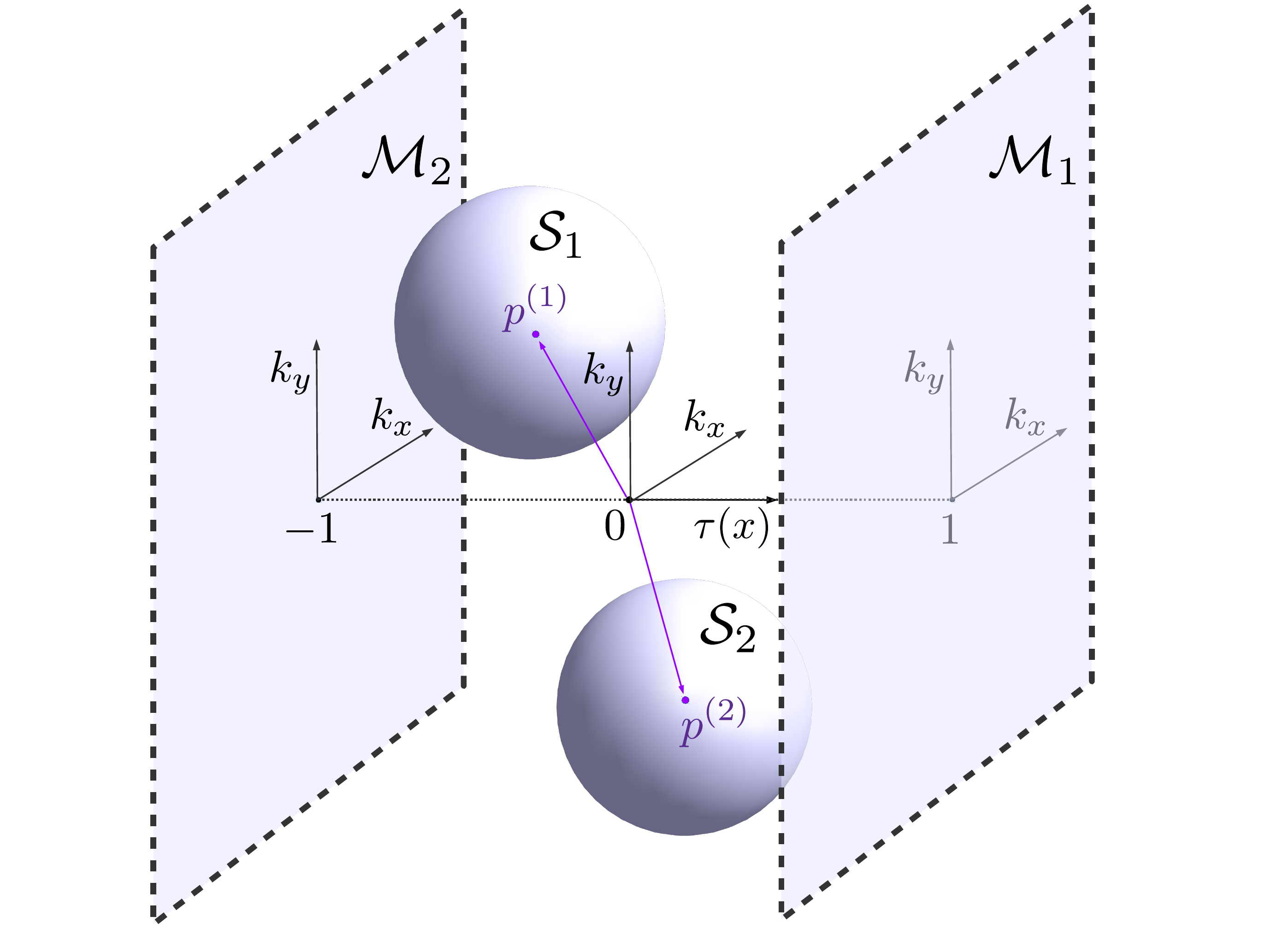}
\caption{Extended base space with the bulk base spaces ${\mathcal M}_{1,2}$ and base spaces ${\mathcal S}_l$ around generic degeneracy-points (particle/hole symmetric point are not shown).} \label{fig:ext_bs}
\end{figure}

  From this symbol we can define the projectors fiber bundles as we did in \cref{sec: top_and_bb} with $L_{\omega,\bf k}$, with the difference that its base space is extended to include the parameter $x$. Since we are going to consider linear interpolations \cref{M_interp} in the examples we will present, we include the interface parameter via the $\tau$ function rather than $x$ itself. This choice will not affect any result. Hence the new base space is made up of terns $(\tau,k_x,k_y)\in [-1,1]\times\mathbb{R}^2$. Next, we identify points $p^{(l)} = (\tau^{(l)},k_x^{(l)},k_y^{(l)}),\;l=1,2,\dots$ of band-degeneracy in this extended fiber bundle, i.e. points at which $\omega^{(i)}\vert_{p^{(l)}} = \omega^{(j)}\vert_{p^{(l)}}$, $i$ (resp. $j$) denoting an upper (lower) band. For the sake of simplicity, here we consider only cases where such points are isolated (Berry monopoles). To proceed further we construct small 2-spheres $\mathcal{S}_l$ that enclose each monopole separately (in the case of line degeneracies we would construct enclosing cylinders \cite{Bra2016}). Finally, we consider the fiber bundles having these spheres as base spaces and the eigenvectors of one of the bands involved as fibers (see \cref{fig:ext_bs}).
  
Assuming there is a gap shared by the right and left bulk, we claim the following spectral flow formula
\begin{equation} \label{bb_corresp2}
    N^{l,\sigma}_{chiral} = - C^{(l,\sigma)}
\end{equation}
where $N^{l,\sigma}_{chiral}$ is the number of localized states in the gap flowing towards the band $\sigma$ as $k_y$ goes from lower to higher values than $k^{(l)}_y$; $C^{(l,\sigma)}$ is the Chern number, as from \cref{ch_num}, with $\mathcal{M}=\mathcal{S}_l$ and $P$ projecting on the eigenvectors of the band $\sigma$.

 The formula is known (in a more abstract fashion) in the context of micro-local (or semi-classical) analysis \cite{Tay1996,Fed1996} and represent a remarkable link between the operatorial spectral flow $N^{l,\sigma}_{chiral}$ and the topological data of its symbols $C^{(l,\sigma)}$. It is strictly valid in the socalled "semi-classical regime", however as we will see later in the examples, this assumption is not always necessary. For more details one can refer to \cref{app:symbol}. As mentioned above the formula has found already application in condensed matter and in different realms, for instance for the description of spectra of molecules with spin-orbit coupling and equatorial waves, where it appear in this exact form\cite{Fau2019,Fau2001}. 

Some additional remarks about the formula are in order:

\textit{i)} The minus sign in the formula depends on the orientation of the axis-tern of extended base space. Here we agree with the convention of Refs.  \cite{Fed1996} and  \cite{Fau2019}.

\textit{ii)}  The formula can be extended to the case where more than two bands are involved in the degeneracy. Whenever this is the case, more than one gap have to be considered, and the spectral flows concern only the gaps between (spectrally) adjacent bands. 

\textit{iii)} There will always be a degeneracy line at $|\omega| = |\bf k| = \infty, \;\forall\,\tau$ due to the vacuum response limit of $M$. However, the projectors $P$ in that limit are the same for any material. This invariance guarantees that no role is played by this degeneracy. 

\textit{iv)}  When $\omega^{(l)}=k^{(l)}_y=0$ and the number of bands involved is odd, particle-hole symmetry enforces the vanishing of $C^{(l,0)}$ for the central band. For the rest, this symmetry does not imply any constraint in the definition of the complex fiber bundles and does not affect the spectral flow formula, the only implication being that for any degeneracy point at $(\tau,k_x,k_y)$ in the extended base space there will be a corresponding one at $(\tau,-k_x,-k_y)$.

\textit{v)} The nature of the degeneracy points in the extended base space might depend on the family of matrices $M(\tau)$ describing the interface. This feature is not present in crystals. There, Chern numbers can be computed in the Brillouin zones $\mathcal{M}_1$ and $\mathcal{M}_2$ of the two materials and their difference must equate the total Berry flux emanated by the monopoles across \textit{any} possible homotopy $M(\tau)$ with $M(\pm1)$ kept fixed. This equivalence cannot be stated in the continuum case, instead classes of topologically equivalent interfaces can be defined. A change of class could be determined by degeneracy points coming from or going to $|\bf k|= \infty$ with $-1<\tau<1$.

\section{Application to gyrotropic continuous media} \label{sec: anal_num}
We show now that the spectral flow formula \cref{bb_corresp2} correctly predicts the number of chiral Maxwell modes for various interfaces between different optical continua. We take under consideration three materials: a perfect electric conductor (a Drude metal with high plasma frequency), a gyro-electric material (a Drude magnetized plasma) and a gyro-magnetic one (ferrite), the last two under a magnetic field $B_z$ oriented along $z$. The first material serves as a "trivial insulator" from the point of view of the topological theory, while the other ones are already known to feature topological physics.
The three response matrices are listed in \cref{tab: materials} (we neglect non-hermitian lossy contributions) \cite{Ero2010}.
\begin{table*}[t!]
\centering
\setcellgapes{2pt}
\makegapedcells
\[
\begin{array}{ | c | c | c | c| } 
\hline
               &\mathrm{Metal} & \mathrm{Ferrite} & \mathrm{Magnetized\;plasma} \\ 
\hline
\epsilon \quad & \left(1 - \frac{\omega_p^2}{\omega^2} \right) \mathbb{1}  &  \mathbb{1} &  \begin{pmatrix}                                                  1 - \frac{\omega_p^2}{\omega^2-\omega^2_c}           &  i\frac{\omega_p^2 \,\omega_c}{\omega(\omega^2-\omega^2_c)}  &  0 \\                                                 -i\frac{\omega_p^2 \,\omega_c}{\omega(\omega^2-\omega^2_c)} &  1 - \frac{\omega_p^2}{\omega^2-\omega^2_c}              &  0 \\                                                      0                        &                       0        &    1 - \omega_p^2/\omega^2     
\end{pmatrix} \Tstrut \Bstrut \\
\hline
\mu &  \mathbb{1}  &
\begin{pmatrix}
1 - \frac{\omega_m \,\omega_{b}}{\omega^2-\omega^2_m}  &  i\frac{\omega \,\omega_b}{\omega^2-\omega^2_m}        &  0 \\ 
-i\frac{\omega \,\omega_b}{\omega^2-\omega^2_m}          &  1 - \frac{\omega_m \,\omega_{b}}{\omega^2-\omega^2_m}  &  0 \\ 
0                                  &                    0                                  &    1   
\end{pmatrix} \Tstrut \Bstrut
& \mathbb{1} \\
\hline
\end{array}
\]
\caption{Dielectric coefficients for the materials under consideration.}\label{tab: materials}
\end{table*} 
There, $\omega_p$ is the plasma frequency of the material, $\omega_m \propto B_z $, $\omega_b \propto S_z$ ($S_z$ being the value of the saturated magnetization) and $\omega_c \propto B_z$ is the cyclotron frequency. Notice that the magnetized plasma response reduces to that of the metal when the magnetic field is switched off.
As previously said, the specific form of these responses implies that TE and TM modes are uncoupled. In the following we will focus only on the interesting modes of each case, namely, TE modes for interfaces with ferrite and TM modes for those with magnetized plasma. 

\subsection{Solvable "Spin models" in gyrotropic materials} \label{spinmodel}
Before addressing the interface problem between these different materials, it is instructive to focus on simple cases which are fully analytically solvable. 

\subsubsection{ Solvable "Spin-$1/2$" problem}
The simplest possible model that illustrates the topological spectral flow theory is provided by the two-dimensional Dirac Hamiltonian with an inhomogeneous mass term. Such Hamiltonians turn out to be ubiquitous in condensed mater; they appear as effective descriptions around two-band crossings such as those encountered in topological phases and graphene-like structures. As already mentioned above, Raghu and Haldane provided such an effective description for gyro-magnetic photonic honeycomb crystals. After linearization in the vicinity of one of the two Dirac points in the Brillouin zone, they obtained an effective Dirac Hamiltonian of the form (Eq (76) of \cite{RaghuPRA})
\begin{align}
H(k_x,k_y,\kappa)=
\begin{pmatrix}
\kappa & k_x -i k_y \\
k_x + ik_y  & -\kappa
\end{pmatrix}    
\end{align}
where $\kappa$ is the Faraday coupling. The two frequency bands $\omega_\pm=\pm\sqrt{k_x^2+k_y^2+\kappa^2}$ reveal a two-fold degeneracy point at the origin of parameter space $(k_x,k_y,\kappa)$. Within our framework, this Hamiltonian constitutes the \textit{symbol} from which the topological properties are inferred. In this simple case, the Chern numbers associated to the fiber bundles of positive and negative frequency, constructed from the family of normalized eigenstates parametrized on a sphere that surrounds this two-fold degeneracy, are $C^{(1,\pm)}=\mp 1$. One can then deduce from the spectral flow theorem, that the inhomogeneous problem
\begin{align}
\hat{H}[-i \partial_x, k_y,\kappa(x)]=
\begin{pmatrix}
\kappa(x) & -i \partial_x -i k_y \\
-i \partial_x + i k_y  & -\kappa(x)
\end{pmatrix}    
\end{align}
where $\kappa(x)$ describes a smooth variation of the Faraday coupling that changes sign with $x$, hosts one chiral mode around $k_y=0$ that leaves the negative frequency band and reaches the positive frequency band, according to $N_{chiral}^{1,\pm}=\pm1$.
\begin{figure}[b!]
\includegraphics[width=.5\textwidth]{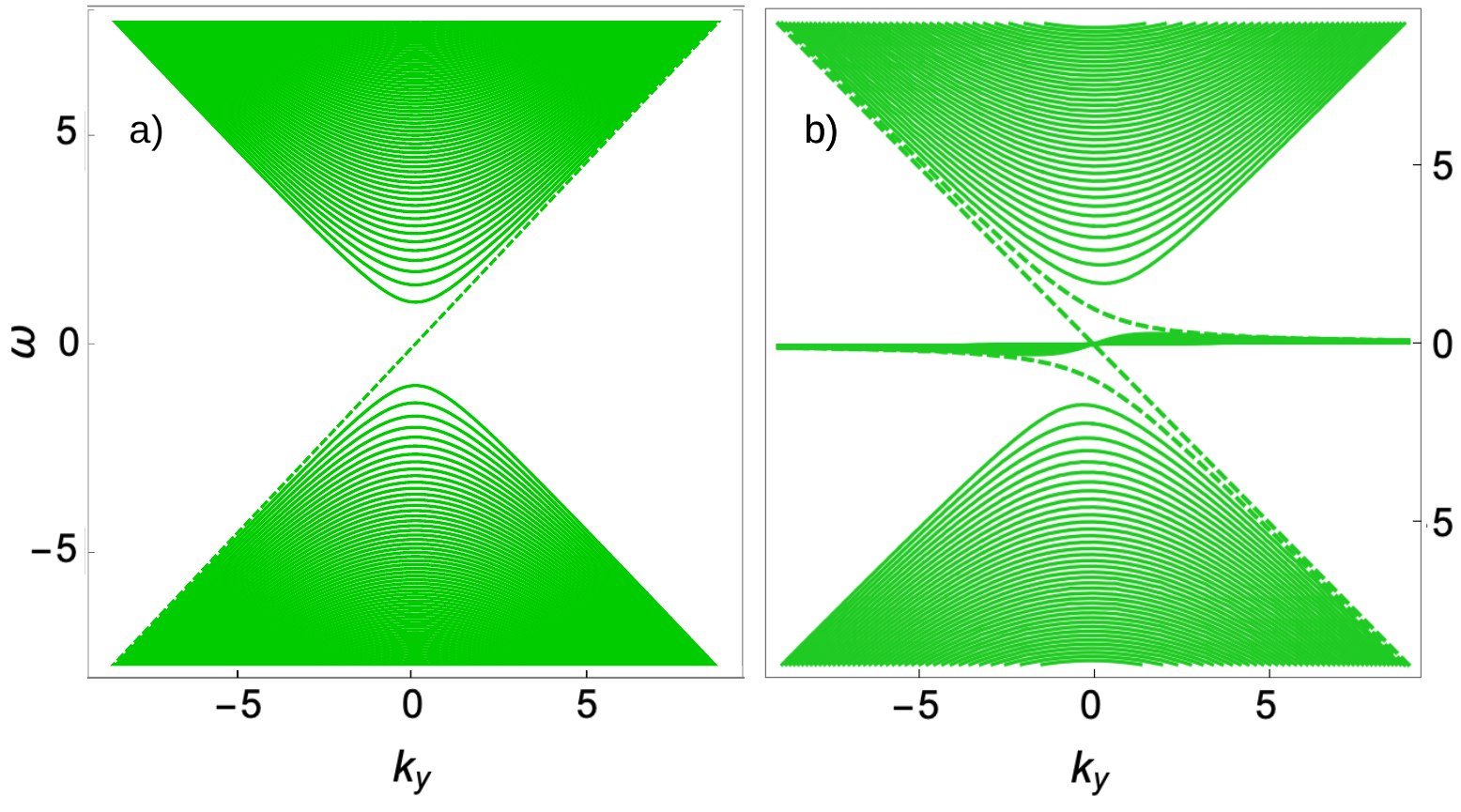}
\caption{ a) Dispersion relation of the linearized Raghu-Haldane model for gyro-magnetic 2D photonic crystal with $\kappa(x)$ a linear function of $x$. b) Dispersion relation of the transverse magnetic modes in a gyro-electric continuum media with $\Omega(x)$ a linear function of $x$.} \label{fig:SW_gyro}
\end{figure}

This spectral flow, that can be simply computed analytically in the case of a linear variation $\kappa(x)\sim x$ (see figure \ref{fig:SW_gyro}(a)), corresponds in that case to an interface mode of the form $e^{-x^2/2}\,(1,i)^T$. This mode was already found by Raghu and Haldane \cite{RaghuPRA}, although the topological description of the system was addressed at the level of gapped bands in the Brillouin zone rather than that of the degeneracy points as we do here. In the next example, we consider in more details a continuum model that is not derived from a photonic crystal.

\subsubsection{Solvable "Spin-1" problem} \label{spin1model}
This model is obtained for the magnetized plasma in the regime $\omega_c \gg \omega_p, \omega$. According to the permittivity tensor in \cref{tab: materials} in that limit, the Maxwell's equations \eqref{Max1} for the TM modes simplify into the explicit eigenfrequency problem
\begin{align} \label{shallow_water}
\begin{pmatrix}
0 & i \Omega & -k_y \\
-i \Omega      & 0  & k_x \\
-k_y  &  k_x  &  0
\end{pmatrix}
\begin{pmatrix}
E_x\\
E_y\\
H_z
\end{pmatrix}
=\omega 
\begin{pmatrix}
E_x\\
E_y\\
H_z
\end{pmatrix}
\end{align}
where $\Omega=\frac{\omega_p^2}{\omega_c}$.

Note that this electromagnetism model is (up to a rotation $k_y \rightarrow -k_x$ and $k_x \rightarrow k_y$) formally equivalent to the linearized two-dimensional shallow water model encountered in fluid dynamics, where the electric field plays the role of the in-plane fluid velocity, and where the magnetic field component plays the role of the thickness variation of the fluid. The fluid mechanics analog of the frequency parameter $\Omega$ is the Coriolis parameter. In the geophysical context, this parameter changes sign at the equator, giving rise to two eastward waves trapped at the equator as found by Matsuno  \cite{Mat1966}. The topological properties of this model have been unveiled recently \cite{Del2017} and here we present its electromagnetic analog.
The eigen-frequencies of this model constitute three continuous bands $\omega_{\pm}=\pm\sqrt{k_x^2+k_y^2+\Omega^2}$ and $\omega_0=0$ that touch in parameter space $(\Omega,k_x,k_y)$ at $p^{(0)}=(0,0,0)$. As explained in the previous section, a fiber bundle can be constructed from each eigenstate on a sphere that encloses this three-fold degeneracy point in the parameter space (see figure \ref{fig:ext_bs}).
In the local spherical coordinates, where $k_x=k \cos \phi$, $k_y=k \sin \phi$ and $\Omega = \omega_+ \cos{\theta}$, the projector on the band of positive frequency reads
\begin{align}
P_+ \!=\! \frac{1}{2}  \!\!\begin{pmatrix}
s^2_\phi + c^2_\theta c^2_\phi & -s^2_\theta s_\phi c_\phi+i c_\theta& s_\theta(i c_\theta c_\phi-s_\phi)\\
-s^2_\theta s_\phi c_\phi-i c_\theta & c^2_\phi +c^2_\theta s^2_\phi & s_\theta(i c_\theta s_\phi + c_\theta)\\
s_\theta(-i c_\theta c_\phi - s_\phi)  & s_\theta(-i c_\theta s_\phi+c_\phi) & s_\theta^2
\end{pmatrix}   
\end{align}
where the shorthands $s$ and $c$ denote sine and cosine functions. 

From formula \eqref{ch_num}, one can finally infer the value of the Chern number for this fiber bundle that is $C^{(1,+)}=2$. 
This topological property guaranties that two chiral modes reaching the positive frequency band emerge for the dual problem where one considers now a smooth spatial variation  of $\Omega$ along $x$ that changes sign,as for instance $\Omega \rightarrow  \Omega(x)= \frac{\omega_p^2}{\omega_c}\, x/x_0$ where $x_0$ is an arbitrary length that we can choose such that $x_0=\omega_c/\omega_p^2$. The eigenvalue problem becomes 
\begin{align}
\begin{pmatrix}
0 & i\, x & -k_y \\
-i \, x      & 0  &  -i \partial_x \\
-k_y  &  -i \partial_x  &  0
\end{pmatrix}
\begin{pmatrix}
E_x\\
E_y\\
H_z
\end{pmatrix}
=\omega 
\begin{pmatrix}
E_x\\
E_y\\
H_z
\end{pmatrix}
\end{align}
and can be solved analytically by use of Hermite polynomials. The frequency spectrum is shown in \cref{fig:SW_gyro}(b). The plots label need to be devided by $k_0$ and $\omega_0$. It shows a spectral flow (dashed lines) in agreement with the topological numbers, where the two, dispersive and non-dispersive, chiral waves have the respective dispersion relations $\omega=\mp\frac{1}{2}(k_y\pm \sqrt{k_y^2+4})$ and $\omega=- k_y$. These modes are trapped around $x=0$ with the following profile along the $x$ direction
\begin{align}
\begin{pmatrix}
E_x\\
E_y\\
H_z
\end{pmatrix}
=
\left\{
    \begin{array}{ll}
        \begin{pmatrix}
e^{-x^2/2}\\
0\\
e^{-x^2/2}
\end{pmatrix} & \text{non dispersive} \\
\begin{pmatrix}
i\frac{x}{\omega-k_y}e^{-x^2/2}\\
-e^{-x^2/2}\\
i\frac{x}{\omega-k_y}e^{-x^2/2}
\end{pmatrix}
& \text{dispersive}
    \end{array}
\right.
\end{align}

\subsection{Beyond spin models: Numerical simulations} \label{sec: num_sim}

We are now ready to move on to more complex situations by performing numerical simulations of the following interfaces $M_1/M_2$: 
\begin{itemize}
\item[(a)] ferrite ($B_z\!\!>\!\!0$)/ferrite ($B_z\!\!<\!\!0$)
\item[(b)] ferrite/metal
\item[(c)] magnetized plasma ($B_z\!\!>\!\!0$)/magn. plasma ($B_z\!\!<\!\!0$)
\item[(d)] magnetized plasma/metal.
\end{itemize}

In order to avoid unwanted issues with physical boundaries we have set periodic boundary conditions along the $x$-axis i.e. $\bf V(x=x_0,k_y) = \bf V(x=-x_0,k_y)$ ($x_0$ is set as the unit of length) closing the systems to what would be a cylindrical geometry if we neglect the $z$-direction. We interpolated according to \cref{M_interp} and chose the function $\tau(x)$ such that one material is in the half-cylinder (in unit $x_0$) $x\in[-0.5,0.5]$ while the other in $x\in [-1,-0.5] \bigcup [0.5,1]$: 
 \begin{equation} \label{tau_x}
 \tau(x) = \tanh{\left[\alpha(|x|/x_0 - 0.5)\right]} \equiv  \tanh{\xi(x)}
 \end{equation}
 where $\alpha$ is a barrier width parameter and $\xi$ is a rescaled displacement from the barriers.
This geometry allows one to simply highlight the opposite propagating direction of the topological modes localized at different interfaces.

\begin{figure*}[htb!]
\includegraphics[width=1.\textwidth]{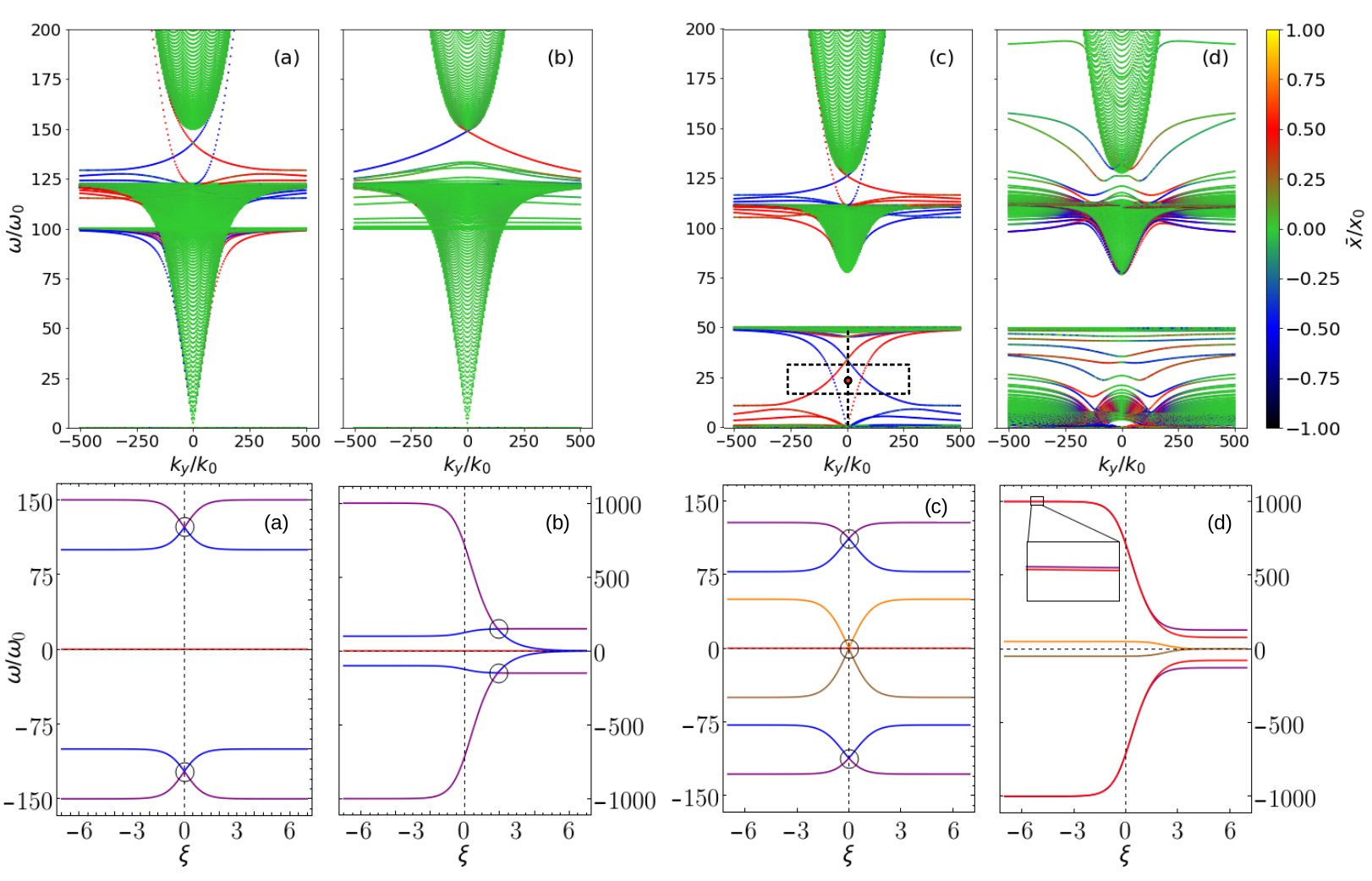}
\caption{ (Upper row) Positive-frequency band dispersions in a cylindrical geometry for different combinations of materials. The colorbar shows the average localization of the eigenstates. The dashed window centered at the mid-gap (red point) in (c) identifies the spectral flow concerning the lower bands. (Lower row) Local eigenfrequencies at $|\bf k|=0$ as a function of the rescaled displacement from the interface. Berry monopoles are highlighted with circles. Cases (a-d) as presented in \cref{sec: num_sim}.
Parameters of the materials (see \cref{tab: materials}): (in unit $\omega_0 = 2\pi c/x_0$) for ferrite, $\omega_b = 50, \omega_m = 100$; for magnetized plasma, $\omega_c = 50, \omega_p = 100$; for metal, $\omega_p = 1000$.} \label{fig:bands}
\end{figure*}

In the upper row of \cref{fig:bands} we show the band dispersions $\omega(k_y) $ computed numerically \cite{Ded2018}. The material parameters for the simulations are written in the caption.  We are not interested in a quantitative description and the parameter values are chosen quite arbitrarily in favor of a good visibility of the topological states. The simulation and interface parameters are common to all cases: $N_x=240$ (number of points for discretization along $x$-axis), $\alpha = 120$. For each instance we show in lower row the eigenfrequencies calculated at $| \bf k|=0$ and highlight the degeneracy points in the extended space in going across the interface at $x= 0.5 \,x_0$.

In these setups TE and TM modes are decoupled. For the cases (a) and (b) we have shown only the TE modes while for cases (c) and (d) only the TM ones. 
Let us comment now on the results in the \cref{fig:bands} case by case:
\begin{itemize}
\item[(a)] There is a single pair of (particle/hole symmetric) Berry monopoles sitting at $(\tau,k_x,k_y) = (0,0,0)$ of Chern number $\left\vert C^{(1,\sigma)} \right\vert=2$ with corresponding frequency $|\omega^{(1)}| \sim 100\omega_0$. 
In agreement with \cref{bb_corresp2}, two chiral states (in yellow in \cref{fig:bands}) traverse the gap at that frequency. Correspondingly, two states flow in the opposite direction at the other interfaces.

\item[(b)] There is still one pair of degeneracy points, however they are shifted at position $\simeq(\tanh{1.9},0,0)$ due to the non-symmetric configuration and $\left\vert C^{(1,\sigma)} \right\vert=1$. A more explicit calculation of the Chern number without using \cref{ch_num} is detailed in  \cref{app: calc_ch_num}.

\item[(c)] This interface is the most interesting. All monopoles are sitting at $(0,0,0)$, but they are of different kind. In particular a pair corresponds to $|\omega^{(1)}| \sim 100\omega_0$ and involves the two uppermost and lowest bands with $\left\vert C^{(1,\sigma)} \right\vert=2$; an other single monopole at frequency $|\omega^{(2)}| = 0$ involves $3$ bands with $C^{(2,\pm)}= \mp2$ for the upper (lower) band,  $C^{(2,0)}= 0$ for the central one. In agreement with \cref{bb_corresp2}, a pair of chiral states traverses the gap at the frequency $|\omega^{(1)}|$ and another pair traverses the gap above $|\omega^{(2)}|$ in the opposite direction. The topological states associated to $|\omega^{(2)}|$ remarkably do not overlap in frequency with the TE modes, the latter lying at higher frequencies, $\omega>\omega_p=100\omega_0$, as explained in \cref{app: TETM}. They are therefore good candidate for a clean experimental detection.

\item[(d)] This interface is a critical one as the monopole is present only asymptotically at $\xi=-\infty$, which is not attained in the simulation. We found no in-gap topological states. Curiously, states emerge outside the bulk bands. They are evanescent states localized at different interfaces but they are hybridized because of the finite size of the $x$-direction. They could be of topological origin \cite{Bli2015} but this analysis goes beyond the purpose of the paper. 
\end{itemize}

 We have checked the convergence of our results by increasing the number of points. While the bulk bands are stable at $N_x = 240$, the in-gap states (topological and non-topological) are not quantitatively converged up to at $N_x = 700$. Nonetheless the topological ones are qualitatively converged at $N_x = 240$, while the non-topological ones are not. The specific cylindrical geometry we have chosen might play a role here. In any case, the latter states do not contribute to the spectral flow, their only influence being to possibly repel the topological states from touching the bulk bands at large $k_y$ (see for instance \cref{fig:bands}(a) ).

A careful look at the lower row of \cref{fig:bands} may reveal an apparent contradiction, already mentioned in Ref. \onlinecite{Sil2016}. The bulk bands (i.e. at $|\xi| >> 1$) of a material may look different in the different subfigures. For instance, consider the metal case at $\xi = -7$ in (b) and (d). In (b) the middle band (in blue) is at $\omega=100\,\omega_0$ while in (d) the corresponding band (in orange) is at $\omega=50\,\omega_0$. These frequencies stem from the frequency-poles of the respective response matrices, that of the ferrite in (b) and that of the magnetized plasma in (d). However, in the limit of infinite $|\xi|$ the residues of these poles vanish implying the disappearance of the electromagnetic mode\footnote{Mathematically, a factor $\omega - \omega_{pole}$ factorizes out in some lines the Maxwell's equations.}. We conclude that their bulk limit is not physically relevant. 
As a check, for cases (a) and (c) we have repeated the simulations with a different interpolation, perhaps more physical, assuming $B_z$ (and therefore $\omega_{m,b,c}$) to be linearly interpolated across the interface. All monopoles are sitting at $(0,0,0)$ with vanishing frequency, we found the sets of Chern numbers to be the same as the cases above. As a consequence the spectral flow is also unchanged. Allowing for $B$ to rotate in the $xz$ or $yz$ planes would allow also to check for the robustness of the spectral flow. However, line degeneracies appears in the extended base space that are left out from the present discussion.
For a comparison of our analysis with an interface problem found in literature see \cref{app:comparison}.

\section{Failure of the standard bulk/boundary correspondence: \newline spin $1$-spin $1/2$ interface} \label{sec: sw_dirac_interf}

In some of the cases presented above, the standard bulk/boundary formula \cref{bb_corresp1} actually works and one might doubt the real need of the spectral flow formula \cref{bb_corresp2} which is sometimes harder to compute. Here we analyze why this happens and show how the spectral flow formula is more general and stable than the other one with an illuminating counterexample.

The materials involved in the previous paragraphs are not topological since the base spaces of their fiber bundles $\mathcal M = \mathbb{R}^2$ cannot be compactified to a sphere. This is a consequence of the fact that the limit of $\bf V(\bf k)$ for $|\bf k| \rightarrow \infty$ exists but depends on the direction of the vector $\bf k$. Nonetheless one can think of a generalization of the formula \cref{bb_corresp1} for these cases. Indeed one can close $\mathcal M$ to a disk $\mathcal{D}$ adding to the bundles the direction dependent limits at infinity of the projectors $P$. 
 Given now an interface problem, if we suppose that the limit at infinity can be defined for all values of the coordinate $x$, then the extended base space would become the cylinder $\mathcal{C}=\{(\tau,k_x,k_y)\,:\,(\tau,k_x,k_y) \in [-1,1]\times\mathcal{D}\}$.

Since the disk has a boundary, $C$ in \cref{ch_num} is not guaranteed to be integer and the standard bulk/boundary formula \cref{bb_corresp1} is of no use. However what is guaranteed to be integer is $C$ computed over the whole boundary of $\mathcal{C}$. This should suggest a straightforward upgrade of the standard bulk/boundary formula where the contribution of the side face of the cylinder $\mathcal C$ is included. 
Surprisingly enough, as mentioned above, this very additional term seams irrelevant in certain cases for it is vanishing. For instance, a close inspection of the analytical model of the magnetized plasma interface in the regime $\omega_c \gg \omega_p, \omega$ presented in \cref{spin1model} reveals that the quantity $C$ in the bulks is actually an integer equal to $\pm 1$ (see \cite{Del2017}) and the formula \cref{bb_corresp1} indeed gives the correct number of interface chiral modes. One may suspect that it is generic that the contribution from the side face is vanishing. 

This suspect is wrong. We present here an enlightening example of a smooth interface between the spin $1$ model of \cref{spin1model} and a fictitious spin $1/2$ material obeying a Dirac continuous equation:
\begin{align}
     m_{1}& = 
\begin{pmatrix}
0 & i \Omega & -k_y \\
-i \Omega      & 0  & -i\partial_x \\
-k_y  &  -i\partial_x  &  0
\end{pmatrix}
\\
     m_{1/2}&  = 
     \begin{pmatrix}
0 & 0 & 0 \\
0  & \Omega  & -i\partial_x + i k_y \\
0  &  -i\partial_x + i k_y  &  -\Omega
\end{pmatrix},
\end{align}
the full interface problem being
\begin{equation} \label{app: interp}
\left[ \frac{m_{1}\! +\! m_{1/2}}{2} + \tau(x)\frac{m_{1} \! - \! m_{1/2}}{2} \right] \bf V = \omega \bf V 
\end{equation}
\begin{figure}[t!]
\centering
\vspace{0.5cm}
\subfloat{%
    \includegraphics[width=0.202\textwidth,trim={.5cm 0 0 -.2cm},clip,valign=t]{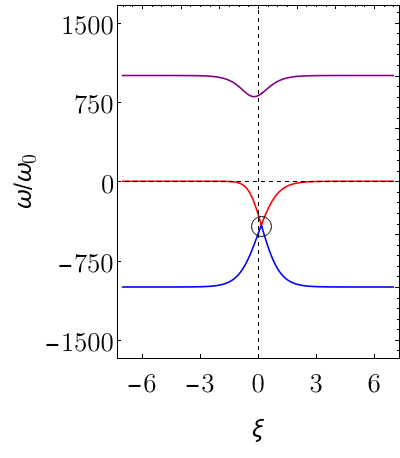}%
  } \quad
  \subfloat{%
    \includegraphics[width=0.26\textwidth,trim={0 0 0 0.2cm},clip,valign=t]{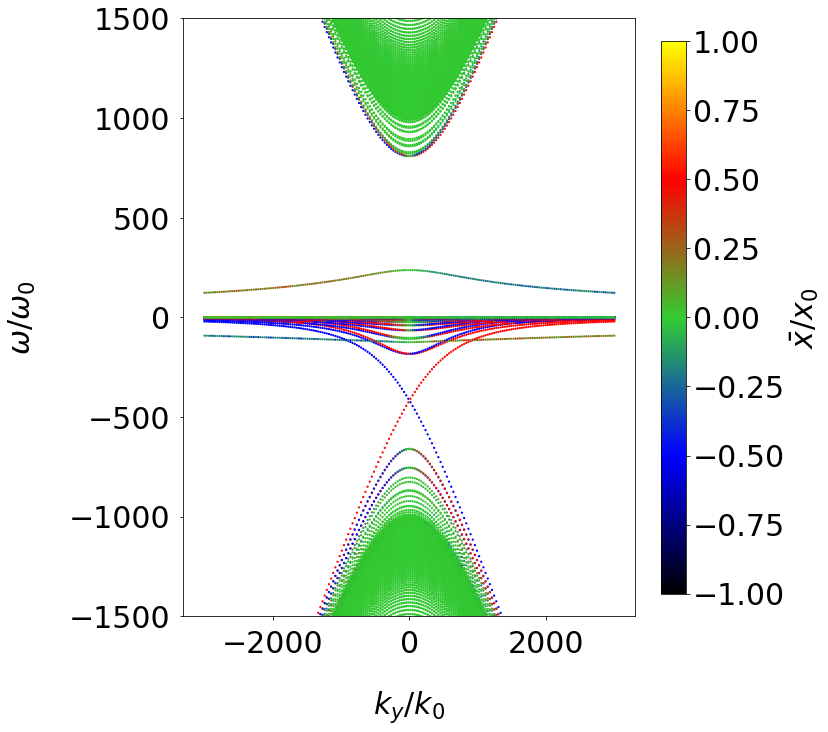}%
  }
  \caption{ Topology from Berry monopoles for an interface between a spin 1 and a spin 1/2 material. (left) Local eigenfrequencies as in \cref{fig:bands}. (right) Eigenfrequency bands in a cylindrical geometry. Parameters (in unit $\omega_0$): $\Omega = 1000,\,N_x=500,\alpha=80$.}\label{fig: shDir}
\end{figure}
where $\tau(x)$ is as in \cref{tau_x}.
\\
Notice that the bulk materials have exactly the same number of bands (three) and spectra. The fact that the Dirac material is not an optical one is of no relevance for the present discussion. 
This example is interesting because the bulk Dirac bands are known to have $|C^{(\pm)}|=1/2$ and $C^{(0)}=0$ while, as said before, the bulk magnetized plasma has $|C^{(\pm)}|=1$ and $C^{(0)}=0$. We might say that the Dirac $\pm$ bands are not topological while  those of the plasma are, if we assume the fibers on the boundary of the $\mathcal{D}$ to be unaffected by possible deformations. For the interface under consideration, the application of \cref{bb_corresp1} would only lead to the meaningless conclusion that the flow onto the $\pm$ bands is half-integer. In \cref{fig: shDir} we show the local eigenfrequencies and the band dispersions of the interface problem in cylindrical geometry. As can be seen from the figure, a Berry monopole (of charge $1$) is present in the extended space and the numerically computed bands feature a pair of chiral states flowing in opposite directions at the two different interfaces of the cylindrical geometry. Notice the absence of particle/hole symmetry in the problem at hand. 
\\
All of this shows that our spectral flow formula is more "fundamental" and powerful than the standard bulk/boundary one. The latter might be corrected adding a term, but only when the limits at infinity of $\bf V(\bf k)$ exist in the extended base space. Therefore, we can claim that interfaces of continuous media can be topological even when the single materials involved are not.

\section{Conclusions} \label{sec: concl}
We showed how topological chiral modes at smooth interfaces between continuous optical media can be predicted by means of a spectral flow formula, even when the materials involved are not topological on their own. In this context Berry monopoles happen to more "fundamental" than band Chern numbers, being the former local and always calculable quantities while latter global ones of limited usability. Our claims are supported by an analytical low-frequency model for gyro-electric TM modes and numerical studies of several interfaces where we predicted the existence of chiral states. This work calls for a number of attractive possible extensions. First, in the paper we have restricted the analysis to spectral flows at vanishing $k_z$. Enabling finite values opens up to the full 3D spectrum and we expect Weyl physics to take place. Second, it would be interesting to consider hybrid interfaces between continuous media and photonic crystal or, conversely, interfaces between hybrid materials that are continuous along some dimensions and patterned along the others. Third, linear degeneracies \cite{Bra2016} have been left out from our work, however they seem to appear quite naturally in the continuous context. Their inclusion will lead to a more complete version of the spectral flow formula. Finally, we have limited our discussion to the class of materials identified by Raghu and Haldane. To investigate more general responses, for instance bearing losses \cite{Zhe2015,Sil2019} or relative to meta-materials \cite{deN2019,Jia2004}, will allow to make contact with more promising and realistic continuous media interfaces in future works.  

\section{Ackwoledgements}
This work was supported by the French Agence Nationale de la Recherche (ANR) under grant Topo-Dyn (ANR-14-ACHN-0031).

\newpage
\appendix
\crefalias{section}{appsec}

\section{Mathematical aspects of the spectral flow formula} \label{app:symbol}
We show here in more details the proof of the spectral flow formula \cref{bb_corresp2} and discuss its technical aspects. We begin introducing the notion of differential and pseudo differential operator and its associated symbol. Then, we show how generic source-free inhomogeneous local and (sufficiently well-behaving) non-local Maxwell's equations are solved by the kernel of an operator of the latter kind. Finally we state two theorems that characterize the spectral flow across spectral gaps and discuss their implementation.   

\subsubsection{Pseudo-differential operators and H\"ormander symbol class}
We provide here the definitions of differential and pseudo-differential operators along with their associated symbols. Particular attention is devoted to the H\"ormander symbol class, relevant for the spectral flow theorem. We will mostly follow Refs. \onlinecite{Tay1996,Zwo2012}.

Any differential operator can be written formally as
\begin{equation}
    \hat p[\bf x, -i\partial_{\bf x}] = \sum_{|\alpha|\le n} a_\alpha(\bf x) \, (-i\partial)^\alpha_{\bf x}
\end{equation}
for some finite $n\in \mathbb{N}$. $\alpha$ is a multi-index running over all degrees of freedom of $\bf x\in \mathbb R^n$. $a_\alpha$ is some sufficiently regular function while $\partial^\alpha_{\bf x}$ stands for a specific product of partial derivation along the coordinates of $\bf x$ (e.g. $\partial^{(2,0,1)}_{\bf x}= \frac{\partial^{2}}{\partial^{2} x_1}\frac{\partial}{\partial x_3}$).

When applied to a function $f(\bf x)$, the whole expression can be expressed in a convenient algebraic form: 
\begin{equation}
    \hat p[\bf x, -i\partial_{\bf x}] f(\bf x) = \frac{1}{(2\pi)^n}\int \mathrm{d} \bf k \; p^{\mathrm st} (\bf x, \bf k)\, f(\bf k)\, e^{i \bf k \cdot \bf x}   
\end{equation}
where 
\begin{equation} \label{symb_DE}
    p^{\mathrm{st}}(\bf x, \bf k) = \sum_{|\alpha|\le n} a_\alpha(\bf x) \, \bf k^{\alpha}
\end{equation}
 $p^{\mathrm{st}}$ is called the symbol of the operator $\hat p$ in standard quantization and is the function on the "classical" phase space that corresponds to the "quantum" operator $p$.
 
The choice of the algebraic expression is not unique. Other well known symbols can be associated to an operator, for instance, via the Wigner transform and the Weyl trasform. The latter, called Weyl quantization in this context, is of particular relevance:
\begin{align}
    \hat p[\bf x, -i \hbar \, \partial_{\bf x}] &f(\bf x) =& \\
    &\frac{1}{(2\pi\hbar)^n} \int \mathrm{d} \bf k \, \mathrm{d} \bf y \; p^{\mathrm{w}} ( \frac{\bf x + \bf y}{2}, \bf k)\, f(\bf y)\, e^{i \frac{\bf k \cdot (\bf x -\bf y)}{_\hbar}}
\end{align}
where we have introduced $0\le\hbar\le1$, in order to be able to perform the semi-classical analysis of equations ($\hbar=1$ correspond to the initial operator at hand).

For later use, we remark that for a given operator $\hat q(\bf x, \partial_{\bf x})$ we can arbitrarily define an other operator $\hat q'_\hbar(\bf x, \hbar\,\partial_{\bf x})$ equal to it and their respective symbols will be generically different.

For sufficiently regular operators and fixed quantization scheme, the correspondence operator-symbol is one-to-one. In particular, we choose to work with the well-behaving H\"ormander symbol class $S^m_{\rho,\delta}$, defined by the existence of $m \in \mathbb R,0\le\rho\le1,0\le\delta\le1$ and numbers $0<C_{\alpha,\beta}<\infty$, for all multi-indices $\alpha,\beta$ such that it holds
\begin{equation}
    \vert \partial^\alpha_{\bf x} \partial^\beta_{\bf k} \,p(\bf x, \bf k)\vert < C_{\alpha,\beta} \langle\bf k\rangle^{m - \rho|\alpha| + \delta|\beta| } 
\end{equation}
where $\langle\bf k\rangle = \sqrt{|\bf k|^2 + 1}$. 

Notice there is a natural inclusion $S^m_{\rho,\delta}\subset S^{m'}_{\rho,\delta}$ for $m<m'$. As an example, $p(x,k)=\langle k\rangle^l$ is in class $S^{l}_{\rho,\delta}$. 

Operators corresponding to H\"ormander symbols are instances of so-called pseudo-differential operators. Pseudo-differential operators generalize differential operators in that their associated symbol is not represented by a truncated series but by an asymptotic one. While differential operators are local, pseudo-differential operators are pseudo-local
\footnote{In optics, usually people refers to "local responses" as those where $M$ does not contains spatial derivatives. In mathematics the notion of local operators is less strict and includes operators with a finite number of spatial derivatives (that in optics are considered "non-local") as their action requires only a infinitesimal neighborhood of the point of the function they are applied to. Pseudo-local operator are "non-local" by all means as their action requires more than just an infinitesimal neighborhood of the point.}
for some intervals of $\rho$ and $\delta$ (e.g. $\rho>0$ and $\delta<1$ for H\"ormander standard quantized operators), meaning that their kernel is not necessarily a delta function in $|\bf x - \bf y|$ but decays fast enough at large values.

\subsubsection{Maxwell's equations and the Raghu-Haldane's insight}
A very broad class of optical systems is actually described by H\"ormander operators. Symbols and operators have a matrix structure but all definitions are straightforwardly generalized to their elements. In the case of homogeneous systems the symbol associated to the Maxwell operator is just $L_{\omega,\bf k}$ of \cref{Max1}. The reader can verify that the elements of the rotor matrix $R_{\bf k}$ are all $S^{1}_{\rho,\delta}$. Homogeneous non-local responses demand a treatment in terms of pseudo-differential operators instead of differential ones. Indeed, the symbols of the response matrix $M$ are asymptotic series in $\bf k$ (with infinite terms). For instance, suppose one has a response whose kernel is has a Gaussian spread 
\begin{align} \label{M_ex}
M_\omega(\bf x - \bf x') = G\,h(\omega)\,e^{-|\bf x - \bf x'|^2/(2\sigma)},
\end{align} 
with $G$ a $6\times 6$ hermitian matrix and $h$ a function.

Then, its associated (standard) symbol has again a Gaussian spread $M(\omega,\bf k) = G\,h(\omega)\exp{(-\sigma|\bf k|^2/2)}$ which is a Schwartz function (i.e rapid decaying) in $\bf k$ and its Taylor expansion has infinite terms. Notice that the symbol elements are in the H\"ormander class  $S^{-\infty}_{\rho,\delta} := \cap_m S^{m}_{\rho,\delta}$.

The connection between topological properties of symbols and their associated spectral flows has been established for problems of the form 
\begin{equation} \label{app: Hamiltonian}
    H[\bf x, -i\partial_{\bf x}]\,\tilde{\bf V} = \omega\, \tilde{\bf V}.
\end{equation}
Up to authors knowledge, these results have not been extended to the "self-consistent" equations like the Maxwell's ones where, roughly said, the operator $H$ itself depends on $\omega$. To bridge this gap we make use of the Raghu-Haldane's insight that, under the conditions of \cref{sec: max_intro}, Maxwell's equations can be mapped to the spectral equation of the form \cref{app: Hamiltonian}. In particular the spectrum of $H$ coincides with the set of frequencies $\omega$ for which the Maxwell's equations have non-trivial solutions 
\footnote{The mapping has been proven for photonic crystals \cite{Rag2008} and for continuous homogeneous systems \cite{Sil2015} but have not been generalized to inhomogeneous cases. We argue that the mapping holds both for the operator $\hat L$ and for its symbol $L$. The proof resembles the one given in the Appendix A and B of Ref. \cite{Sil2015}. Working with the symbol, the main difference with that proof is that the poles in $\omega$ of the matrix $M$ are functions of both $\bf k$ and $x$. Working with the operators, all matrices in the proof can be replaced by operator; the existence of square roots of positive (possibly unbounded) operators is stated in A. Wouk, SIAM Review \textbf{8}, 1, 100-102 (1966).}.
The only existence of this mapping will justify our strategy of obtaining the spectral formula \cref{bb_corresp2} directly from \cref{Max1} without making an explicit use of the mapping.

\subsubsection{The spectral flow theorem}

\begin{figure}[t!]
\hspace*{-.2in}
\includegraphics[width=0.55\textwidth]{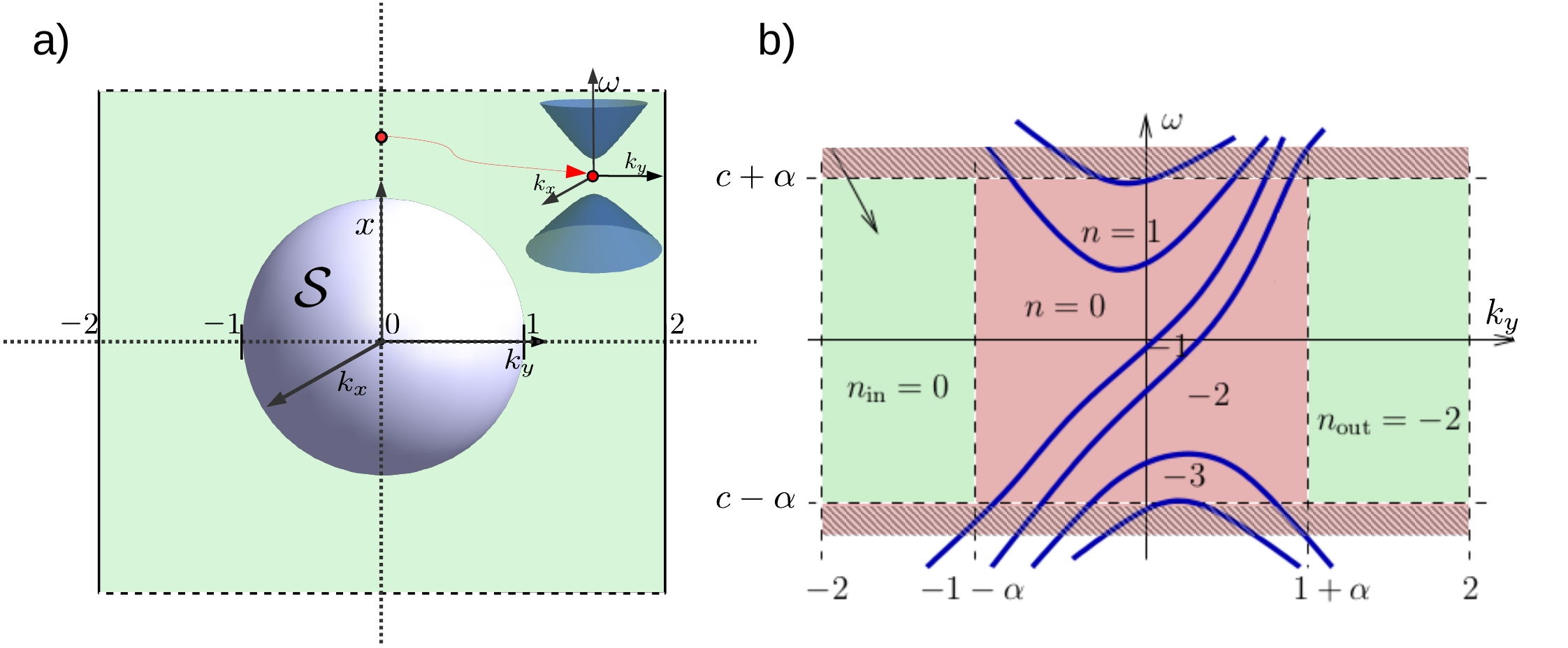}
\caption{Illustration of \cref{theo_1}. a) Hypotesis: gapped region of the symbol in the extended base-space (green area), 3-disk possibly enclosing a band-degeneracy inside the are $\mathcal S$ (in violet), sketch of the gapped bands of the symbol (in blue) at a point in the green area. b) Claim: no spectrum in the green area, discrete spectrum in the pink area. Adapted from Ref. \onlinecite{Fau2019}.} \label{fig:faure}
\end{figure}

There are two theorems about spectral flows of Hermitian operators acting on $\mathbb R^2$ that are of interest (Theorem 2.2 and Theorem 2.7 in Ref. \onlinecite{Fau2019}). We limit ourselves to write only their statements. The first theorem is the following: 
\begin{theorem} (\textbf{Discreteness of in-gap spectrum}) \label{theo_1} 
Let $H_{k_y}(x,k_x)$ be a family Hermitian $N\times N$ H\"ormander symbols whose eigenvalues, labeled by $j$, are $\{\omega_j (x,k_x,k_y)\}$. Suppose there exists an index $r$ and $c>0$ such that $\omega_r (x,k_x,k_y)<-c$ and $\omega_{r+1} (x,k_x,k_y)|>c$ if $||(x,k_x,k_y)||\ge1$ and $|k_y|\le2$ . \\
Then $\forall \alpha>0\;\exists\,\hbar_0$ such that $\forall\,\hbar<\hbar_0$ the Weyl quantized operators $\hat H_{k_y}(x,\hbar\,\partial_x)$ have: \\
i)  discrete spectrum in $(-c + \alpha, c - \alpha)$ that depends continuously in $\hbar$ and $k_y$,  for $|k_y|<1-\alpha$;  \\
ii) no spectrum in $(-c + \alpha, c - \alpha)$ for $1+ \alpha <|k_y|<2$
\end{theorem}

We comment briefly on the theorem. The family of Hermitian operators are the Raghu-Haldane Hamiltonians obtained via the mapping from the Maxwell operators (with $k_y$ as external parameter). Roughly speaking, the hypotheses of the theorem consist in singling out a unit 3-disk of the extended base space $(x,k_x,k_y)$ (inside which two bands might be degenerate) outside which the spectrum is gapped (see \cref{fig:faure}a). The precise position and radius of the disk (set respectively to the origin and $1$), the strip width in $|k_y|$ (set to $2$), and the position of the eigenvalues $\omega_r$ (centered around $0$) are not relevant as the Hamiltonians can be multiplied by a scalar and shifted with an identity matrix, the extended base space can be rescaled as well. For what concerns us the theorem states that inside the shared spectral gap of the bulk materials making up the interface there might be some spectrum due to the interface and \emph{it must be discrete} and localized in $k_y$ (see \cref{fig:faure}b). In particular this is true if the quantization of the symbol is sufficiently small i.e. $\hbar<\hbar_0$. The discreteness of the spectrum inside the gap allows to define the so-called spectral flow:
\begin{equation} \label{N_chir_def}
    N_{chiral} = \#\{\omega<0 \;\mathrm{at}\; k_y=+2\} - \#\{\omega<0  \;\mathrm{at}\; k_y=-2\} 
\end{equation}
Notice that $N_{chiral}$ is independent of $\hbar$, because of the continuity in $\hbar$ of the discrete spectrum, and in general of continuous deformation of the family of symbols. It is therefore a topological index.  

The precise statement is the content of second theorem: 
\begin{theorem} (\textbf{Spectral flow theorem})
Under the same assumptions of \cref{theo_1}, let $\mathcal S$ be the 2-sphere $||(x,k_x,k_y)||\ge1$ and consider the fiber bundle with the projectors to the $r+1$-th band as fibers. If $C$ is the fist Chern number of such bundle and $N_{chiral}$ is defined as in \cref{N_chir_def}, then \\
\begin{equation}
    N_{chiral} = - C
\end{equation}
\end{theorem}
The theorem can be applied to any degeneracy point in the extended base space, around which we can construct the $2$-sphere, as described in the main text. As a corollary it implies that there will be no spectral flow if the sphere does not contain any band degeneracy
\footnote{The sphere $\mathcal S$ in $\mathbb R^3$ is homotopic equivalent to two superimposed disks glued along their boundaries. Therefore the fiber bundle over $\mathcal S$ is homotopically equivalent to a bundle on the disks if there is no band-degeneracy. This implies that the Chern number is vanishing since it is an integral over the two identical domains oriented in opposite directions.}. 

A final remark should be done. In principle we can make use of the above theorems only if $\hbar_0\ge1$. This is usually the case if the operators are not too wild. In any case, one can use a rescaling trick and equate the Maxwell operator $\hat L_{k_y}[\bf x, \partial_{\bf x}]$ to an other operator $\hat L'_{k_y}[\bf x, \hbar' \,\partial_{\bf x}]$ and reconsider \cref{theo_1} with the family of symbols associated to the new operator $\hat L'$ instead of that of $\hat L$. This kind of rescaling will validate the theorem if the new $\hbar'_0<\hbar'$, which is expected to happen if $\hbar'$ is chosen to be small enough.

\section{Calculation of the Chern number} \label{app: calc_ch_num}

The definition of Chern number \cref{ch_num} makes use of the Levi-Civit\'a connection. Our formula is equivalent and have to be compared with the one proposed by Raghu and Haldane for optical systems and widely used in literature \cite{Rag2008}:
\begin{equation} \label{app: ch_num}
    C = \frac{1}{2\pi} \,\int_\mathcal{M} \mathrm{d} c_1 \, \mathrm{d} c_2  \, \left( \partial_1  {\mathcal A_2} - \partial_2  {\mathcal A_1} \right)
\end{equation}
where the (non-standard) Berry connection is ${\mathcal A_j} =  - \mathrm{Im} \, \bf V^\dagger \, \left(\partial_{\omega} \omega M\right)  \partial_j \bf V$. Notice that sometimes ${\mathcal A_j}$ is defined with a minus sign affecting the sign in \cref{ch_num} as well.

The reason for the equivalence of the two expression is that Chern characteristic classes and therefore the Chern number are independent from the specific connection one uses\cite{Fed1996}. A direct consequence is that the polarization modes that contributes to \cref{app: ch_num} through the term $\left(\partial_{\omega} \omega M\right)$ are actually of no topological relevance. In practical calculation it suffice therefore to integrate $P$ in the spherical coordinates of $\mathcal S_l$ with a flat connection i.e. with $(c_1,c_2)=(\theta,\phi)$ and no additional Jacobians in the integrand.
\begin{figure}[b!]
\vspace*{-1.in}
\hspace*{-.5in}
\includegraphics[width=0.55\textwidth]{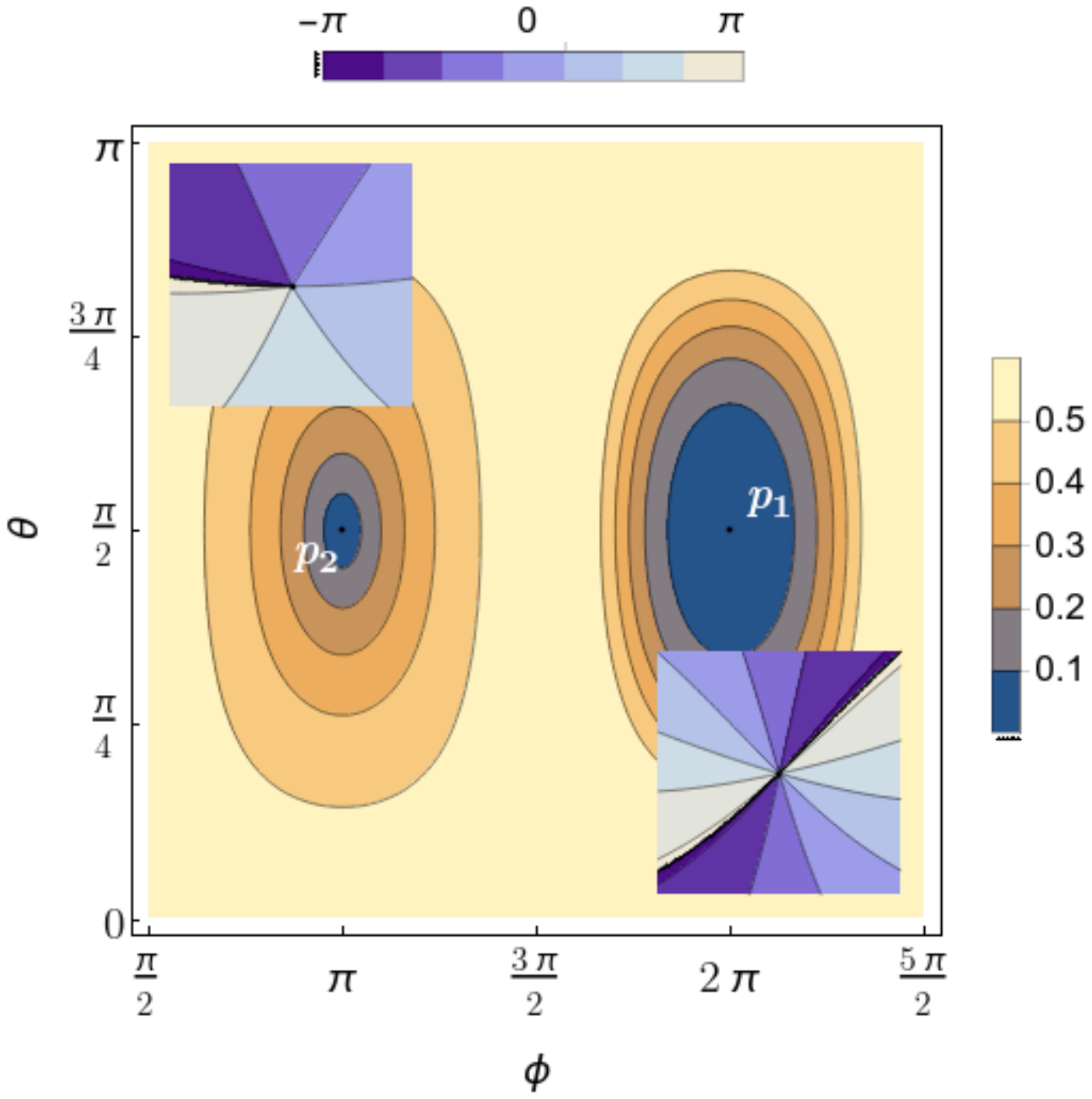}
\vspace*{-1.5in}
\caption{Norm of the global section $s$ over the base space $\mathcal{S}$ parametrized by (shifted) sperical coordinates $(\theta,\pi)$ for the uppermost band of the case ferrite/metal. In the inset, the complex phase of $z_j$ in the vicinity of the vanishing-norm points.} \label{fig:globsec}
\end{figure}
To verify the correctness of the results we have also calculated $C$ by means of a different technique explained in Ref.  \onlinecite{Fau2019}. We consider the projector $P$ onto a band and construct $s=P\,\bf v$, with $\bf v$ an arbitrary vector. $s$ is a global section of the vector bundle on $\mathcal{M}$ that will vanish generically at some set of points $\{ p_j \}$. Then we consider the winding of the complex phase of $z_j=s^\dagger\, \bf V_{p_j}$ (remember that $P = \bf V \,\bf V^\dagger$) with $s$ evaluated along a path that encircle $p_j$. One can show that $C=\sum_j wind\,(z_j)$. We show the numerical technique applied to the case metal/ferrite as in \cref{fig:bands}(b). The Berry monopole is at $\xi\simeq 1.9$ and $|\bf k| = 0$. Therefore we set the base space $\mathcal S$ at the degeneracy point, a sphere with radius $0.3 x_0$ stretched by a factor $1000$ along the momenta directions. As one can see in \cref{fig:globsec}, the global section related to the uppermost band vanishes in two points $p_1$ and $p_2$ and their associated winding is, respectively, $2$ and $-1$. Therefore the total winding is $1=C$ as indeed confirmed by applying \cref{ch_num} to the \cref{fig:bands}(b).    

\section{TE and TM modes overlapping} \label{app: TETM}

\begin{figure}[t!]
\includegraphics[width=0.5\textwidth]{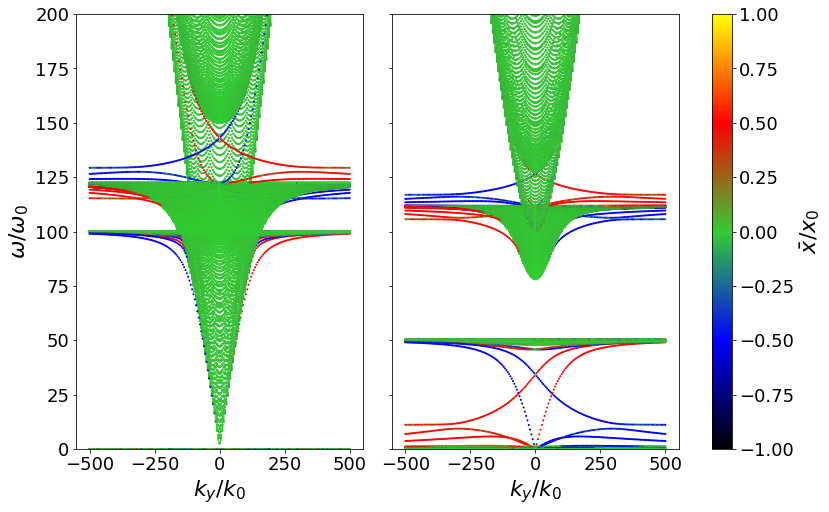}
  \caption{Band dispersion of cases (a) (in the left) and (c) (in the right) of \cref{fig:bands} in the main text. TE and TM modes are shown together.}\label{fig: TETM}
\end{figure}

As mentioned in the main text TE- and TM-modes bands overlap in the band dispersion. In \cref{fig: TETM} we plot again cases (a) and (c) of \cref{fig:bands}, showing together both kind of bands. In the ferrite interface the TM modes have the free-light conical dispersion, therefore they span all energies, in particular also the band gap at $\omega \simeq 140\omega_0$. Since TE and TM modes are coupled in usual experiments due to impurities or boundary effects, we cannot expect chiral edge states to be easily detected.  Fortunately, this inconvenient is not universal. The TE-modes bands dispersion of the magnetized plasma case is actually gapped for $0<\omega<\omega_p$ since the dispersion is that of a Drude plasma. As a consequence the spectral flow of the topological chiral TM modes sitting in the lower gap is robust under disorder and boundary effects.  

\section{Comparison with Ref.  \onlinecite{Sil2016}} \label{app:comparison}

\begin{figure}[t!]
\centering
\vspace{0.5cm}
\subfloat{%
    \includegraphics[width=0.202\textwidth,trim={.5cm 0 0 -0.3cm},clip,valign=t]{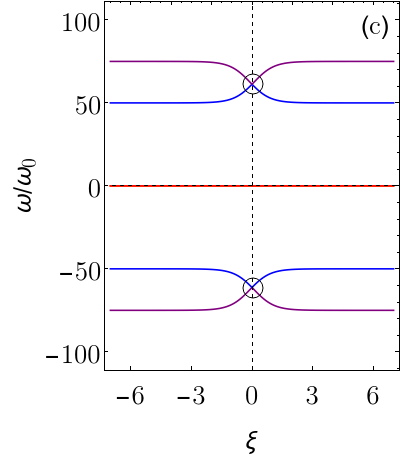}%
  } \quad
  \subfloat{%
    \includegraphics[width=0.26\textwidth,trim={0 0 0 0.2cm},clip,valign=t]{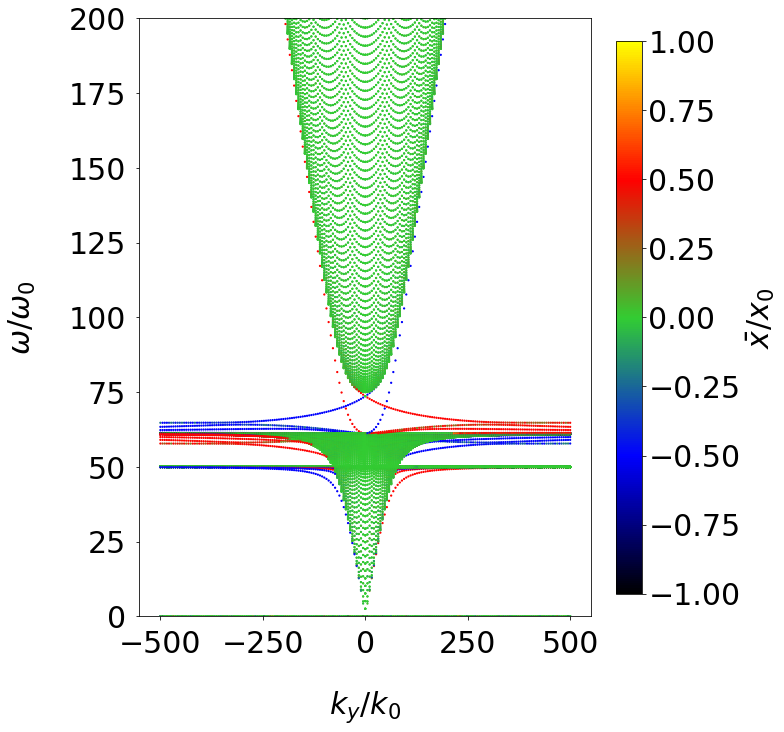}%
  }
  \caption{ Topology from Berry monopoles for a the gyrotropic material of Ref.  \onlinecite{Sil2016}. (left) Local eigenfrequencies as in \cref{fig:bands}. (right) Poisitive-frequency bands in a cylindrical geometry. Parameters (in unit $\omega_0$): $\omega_m = 50,\,\omega_b=25$.}\label{fig:silv}
\end{figure}

We apply our approach to a case studied in Ref.  \onlinecite{Sil2016}. 
The author considers a gyrotropic material which has the same response matrix as that of ferrite if we would have inverted permittivity and permeability: $\epsilon=\mu^{ferr}$ and $\mu=\epsilon^{ferr}$ (cf. \cref{tab: materials}). As mentioned in the main text, in order to compactify the base space a momentum cutoff in front of the gyrotropic part of the response is added such that the off-diagonal terms vanish at big $\bf k$. The spectral flow of a system with a sharp interface between the material with opposite magnetization at opposite sides of the interface is considered. With the cutoff the band Chern numbers are well defined and, using \cref{bb_corresp1}, $N^+_{chiral}=2$ is found for the uppermost band. It is of our interest the situation with no cutoff inserted where the band Chern numbers cannot be computed. The author finds only a single edge states, the missing one being "flown" towards infinite momenta. We have reproduced the result using the parameters of Fig. 8 in the article. For our theory to be applicable we assume a smooth interface. As it can be seen from our \cref{fig:silv}, we have a single pair of Berry monopoles of charge $|C|=2$ therefore our result agrees with the finding of the reference at finite cutoff. In agreement with our theory we find two chiral states traversing the gap instead of one as expected for a sharp interface according to the reference. Additional (not shown) numerics indicates that, as the interface becomes sharper, one topological in-gap mode (the one that extended in $k_y$ the most) increases in energy in agreement with the result of Ref. \onlinecite{Sil2016}.

\end{document}